\documentstyle[aps,prl,floats,epsf]{revtex}

\def\th{\theta}

\def\beq{\begin{equation}}\def\eeq{\end{equation}}
\def\beqa{\begin{eqnarray}}\def\eeqa{\end{eqnarray}}
\def\barr{\begin{array}}\def\earr{\end{array}}

 \let\br=\bigr
 
\let\bm=\bibitem

\def\bd{\begin{document}}
\def\ed{\end{document}}
\def\ba{\begin{array}}
\def\ea{\end{array}}
\def\bea{\begin{eqnarray}}
\def\eea{\end{eqnarray}}
\def\ft#1#2{{\textstyle{{\scriptstyle #1}\over {\scriptstyle #2}}}}
\def\fft#1#2{{#1 \over #2}}
\newcommand{\be}{\begin{equation}}
\newcommand{\ee}{\end{equation}}
\newcommand{\eq}[1]{(\ref{#1})}
\def\eqs#1#2{(\ref{#1}-\ref{#2})}
\def\det{{\rm det\,}}
\def\tr{{\rm tr}}
\newcommand{\ho}[1]{$\, ^{#1}$}
\newcommand{\hoch}[1]{$\, ^{#1}$}
\def\ra{\rightarrow}

\def\Xh{\hat{X}}
\def\ah{\hat{a}}
\def\xh{\hat{x}}
\def\yh{\hat{y}}
\def\ph{\hat{p}}
\def\G{{\cal G}}
\def\Dth{{\Delta_\th}}

\def\bk{{\bf k}}
\def\bx{{\bf x}}
\def\br{{\bf r}}

\begin{document}

\twocolumn[\hsize\textwidth\columnwidth\hsize\csname@twocolumnfalse\endcsname

\title{Inflation as a Probe of Short Distance Physics}
 
\author{Richard Easther$^1$, Brian R. Greene$^{1,2}$, William
H. Kinney$^1$, Gary Shiu $^3$} 
\address{$^1$ Institute for Strings, Cosmology and Astroparticle
  Physics, Columbia University, New York, NY 10027, USA} 
\address{$^2$ Department of Mathematics, Columbia University, New York,
NY 10027,   USA} 
\address{$^3$ Department of Physics and Astronomy, University
  of Pennsylvania,
Philadelphia, PA 19104, USA 
\\ \mbox{}\\Email:     easther@phys.columbia.edu, greene@phys.columbia.edu, 
kinney@phys.columbia.edu, shiu@dept.physics.upenn.edu}

\maketitle
 
{\tighten
\begin{abstract}
 
We show that a string-inspired Planck scale modification of general
relativity can have observable cosmological effects. Specifically, we
present a complete analysis of the inflationary perturbation spectrum
produced by a phenomenological Lagrangian that has a standard form on
large scales but incorporates a string-inspired short distance cutoff,
and find a deviation from the standard result. We use the de Sitter
calculation as the basis of a qualitative analysis of other
inflationary backgrounds, arguing that in these cases the cutoff could
have a more pronounced effect, changing the shape of the spectrum.
Moreover, the computational approach developed here can be used to
provide unambiguous calculations of the perturbation spectrum in other
heuristic models that modify trans-Planckian physics and thereby
determine their impact on the inflationary perturbation spectrum.
Finally, we argue that this model may provide an exception to
constraints, recently proposed by Tanaka and Starobinsky, on the
ability of Planck-scale physics to modify the cosmological spectrum.

\end{abstract}
\pacs{PACS numbers:  75.10.Jm, 67.40.Db, 05.30.Pr }
}
]\narrowtext 

\section{Introduction}

In the last two decades, particle physics has become an indispensable
part of cosmology. In fact, one of the strong
motivations for studying particle physics theories that go beyond the
standard model and incorporate gravity is that they may shed light on
the nature of the cosmological singularities arising in general
relativity.  There is widespread hope that in one form or another
these and other cosmological considerations may one day allow us to
test physical theories whose fundamental scales are now, and perhaps
forever, beyond the reach of conventional accelerator experiments.

Inflationary cosmological models, in particular, significantly
highlight the important role of microphysics. For example, since the
pioneering work of \cite{StructureFormation} it has been known that
quantum field fluctuations in the early universe are stretched by
inflationary expansion to scales of astrophysical relevance, providing
a gratifying first principles mechanism for structure
formation. Galaxies, from this viewpoint, are quantum fluctuations
writ large.

In order to solve the standard cosmological puzzles, a minimum of 60
e-folds of inflationary expansion must be invoked, but in many models
this number can be {\it much} larger. Taking this at face value, it
means that today's scales of cosmological relevance expanded from
Planckian or sub-Planckian scales at the onset of inflation. Inflation
may therefore provide a kind of Planck scale microscope, stretching
the smallest of distance scales to observably large size.

It is possible, however, that in the process of such inflationary
expansion the effect of Planck scale physics gets washed out, being
diluted by the very growth of scales which potentially makes them
visible. A similar phenomenon has been observed in black hole
physics. If one traces the history of a Hawking radiated photon, one
finds that it gets ever more blue-shifted toward the moment of its
emission, and hence one might think it could carry an imprint of
extremely high energy physical processes. In reality though, a number
of studies \cite{BlackHoles} have concluded that such short distance
physics does not have any impact on the low energy features of Hawking
radiation; heuristically, short distance modifications are washed out 
by the memoryless process of thermalization.

In a cosmological context, the situation in this regard has been less
clear.  Brandenberger and Martin \cite{BM} carried out a study of the
impact of various models of trans-Planckian physics (in fact, the
same models considered in the black hole studies just mentioned) on
the spectrum of density perturbations in power law inflationary
models. They found that while some hypothesized Planck scale
modifications to ordinary field theory yield no late time consequences
(similar to the black hole conclusion) some do, indicating a
cosmological sensitivity to short distance physics. 
On the other hand, it was argued in \cite{N} that there should be
no change in the perturbation spectrum if the proposed modifications still
yield the adiabatic vacuum\footnote{Note that the definitions of 
adiabaticity used in \cite{BM} and \cite{N} 
are slightly different, leaving the situation somewhat ambiguous.}.

In this paper we take up the issue of cosmological sensitivity to short
distance physics, but from a different approach. Namely, rather than
considering the {\it ad hoc} short scale modifications studied in
\cite{BM,N}, we focus on an interesting model introduced in
\cite{Kempf}, whose short
distance physics is designed to incorporate a minimum length, a
development naturally inspired by string theory. 
We present a full analysis of the perturbation spectrum
produced by the presence of such a minimum length in a de Sitter
background. As with {\em any\/} calculation of the inflationary
perturbation spectrum, the modes are normalized well inside the
horizon, and the calculational task is to track their evolution until
they are far outside the horizon, as the perturbation spectrum is 
fully determined by their asymptotic values.  We solve the mode
equation numerically to avoid invoking any new analytical
approximations, and fix the initial conditions for the numerical
solution by matching to an approximate analytical solution of the mode
equations that is valid at very early times.

While completing this work, Kempf and Niemeyer \cite{KN} posted a 
complementary  calculation
of the spectrum in the presence of a short distance cutoff. We contrast
our approach with theirs, showing that the numerical evaluation of the
mode functions leads to a complete description of the spectrum.  We
also expect that the  approach used here will generalize to the
broader problem of determining the impact of trans-Planckian physics
on the inflationary perturbation spectrum, which can manifest itself
as a change to the dispersion relation \cite{BM}, or the initial
normalization of the modes or choice of vacuum.

In what follows, we briefly review the construction of a model with a
minimum length, and discuss the analytical properties of the
perturbation equation. In particular, following \cite{Kempf} we show
that since there is a minimum length, each mode is ``created'' at a
finite time in the past, defined by the moment when its physical
wavelength first exceeds the minimum length. We then show how to
extract an approximate analytical solution that holds near this early
time. This allows us to ``match'' this solution to a numerical
solution of the full mode equation, and compute the amplitude of the
perturbation spectrum as a function of the minimum length. We find
that, indeed, there is an imprint of the short distance cutoff on the
perturbation spectrum. We also confirm that the scenario we study
does not violate bounds coming from late time particle production
and quantum mechanical self consistency. 
More generally, we emphasize that the
effect of any short distance modification can be encoded
in two wavelength dependent functions, providing a phenomenological
approach for systematically analysing deviations from
the standard prediction of the power spectrum. 
We conclude with a brief discussion of the
spectrum we might expect in a more general spacetime, and a summary of
our conclusions. 

We stress at the outset that our intent is to show explicitly that a
modification to conventional physics at the Planck scale can have an
observable effect on precision cosmological measurements. While of
interest in its own right, the model \cite{Kempf} on which we focus
should be viewed as one concrete example in which such calculations
can be reliably performed, allowing us to establish definitively that
cosmological observations may be a window onto Planck scale physics.

\section{Perturbations With a Minimum Length}
  
Motivated by the stringy uncertainty principle \cite{sur}, a
phenomenological Lagrangian which incorporates a short distance cutoff
was recently proposed by Kempf \cite{Kempf}.  In this approach, the
short distance cutoff is modeled by modifying the usual commutation
relation to
\begin{equation}\label{uncertain}
\left[ {\bf x}, {\bf p} \right] = i \hbar \left( 1 + \beta {\bf p}^2
\right).
\end{equation}
The parameter $\beta$ is related to the minimum distance $\Delta
x_{min}$ by $\Delta x_{min} \sim \sqrt{\beta}$. A Lagrangian suggested
by Eq.(\ref{uncertain}) was discussed in \cite{Kempf}.  In this model,
the tensor mode $u_{\tilde{k}}$ obeys the following equation of
motion:
\begin{equation}\label{u_kmode}
u_{\tilde{k}}^{\prime\prime} + \frac{\nu^\prime}{\nu}
u_{\tilde{k}}^\prime +\left(\mu 
 -  \frac{a^{\prime \prime}}{a} 
-\frac{a^\prime}{a} \frac{ \nu^\prime}{\nu}
\right) u_{\tilde{k}} = 0,
\label{uequation}
\end{equation}
where $a$ denotes the scale factor and the prime denotes differentiation
with respect to conformal time $\eta$. Our $u_{\tilde k}$ is equal to
$a^2 \phi_{\tilde{k}}$ from \cite{Kempf}, while $\tilde{k}^i = a
\rho^i e^{-\beta \rho^2/2}$ where $\rho^i$ is the Fourier transform of
the physical coordinates $x^i$, and
\be
\mu(\eta,\rho)  \equiv  
\frac{a^2 \rho^2}{(1-\beta \rho^2)^2},\qquad
\nu(\eta,\rho)  \equiv  
\frac{e^{\frac{3}{2}\beta \rho^2}}{
 \left( 1 - \beta \rho^2 \right)}.
\ee
The cutoff is defined by requiring that $\rho^2 \leq 1 / \beta$, and is
motivated by the minimum distance in string theory. Tensor
perturbations with different comoving wavenumber $k$ reach the cutoff
$\rho^2 = 1/\beta$ at different conformal time $\eta_k$, where $a^2
(\eta_k) = \beta k^2$. For de Sitter space, $\eta_k = -1/\sqrt{e \beta
H^2 \tilde{k}^2}$.

Note that in evaluating the derivatives of $u_{\tilde{k}}$ with
respect to $\eta$, we are holding $\tilde{k}$ (and not the usual
comoving momentum $k$) fixed with time.  It is therefore convenient to
express $\mu$ and $\nu$ in terms of $\tilde{k}$ by introducing the
product-log, or Lambert $W$ function \cite{Lambert}, which is defined
so that $W(xe^x)=x$:
\begin{equation}
\mu = -\frac{a^2}{\beta} \frac{W(z)}{\left( 1+W(z) \right)^2}~,
\quad
\frac{\nu^{\prime}}{\nu} = 
\frac{a^{\prime}}{a} \frac{W(z) \left(5+3 W(z \right))}{(1+W(z))^2}  ~.
\end{equation}
where $z=-\beta \tilde{k}^2/a^2$.  The $W$ function has an essential
singularity when its argument is equal to $-1/e$, and this corresponds
to the precise moment (for a given $k$) when $\eta=\eta_k$. Let us
examine in more detail the nature of the singularity in the equation
of motion at $\eta=\eta_k$.  The Lambert $W$ function $W(z)$ has a
series expansion near the branch point $z=-1/e$ \cite{Lambert}:
\begin{equation}
W(z) = -1 + p -\frac{1}{3} p^2 + \frac{11}{72} p^3 + \dots
\end{equation}
where $p=\sqrt{2(ez+1)}$. The series converges for $|p|<\sqrt{2}$.
The singular point at $\eta=\eta_k$ is irregular because the
coefficients of $u_{\tilde{k}}^{\prime}$ and $u_{\tilde{k}}$ are not
analytic in $\eta-\eta_k$. However, the non-analytic piece is less
singular than $1/(\eta - \eta_k)$ and is therefore subdominant. We can
solve for the leading behavior of $u_{\tilde{k}}$ by extracting the
most singular terms of the equation of motion.  First, write $\eta =
\eta_k (1-y)$, so that the leading terms in the equation of motion give
\begin{equation}\label{u_kmode_y}
\ddot{u}_{\tilde{k}}
- \frac{1}{2y} \dot{u}_{\tilde{k}}
+ \left( \frac{1}{2 \beta H^2} + 1 \right) 
\frac{1}{2y} u_{\tilde{k}}
\end{equation}
where dot denotes derivatives with respect to $y$.  Written in this
form, the $\dot{u}_{\tilde{k}}$ and $u_{\tilde{k}}$ terms both have
divergences that scale as $1/(\eta-\eta_k)$, so the equation for the
leading behavior of $u_{\tilde{k}}$ actually has a {\it regular}
singular point at $\eta_k$.

Since any second order differential equation has a power-law expansion
about a regular singular point \cite{BoyceDiPrima}, we can explicitly
construct the two independent solutions of Eq.(\ref{u_kmode_y}) in the
vicinity of $\eta_k$.  Let $u_{\tilde{k}}$ take the form:
\begin{equation}
u_{\tilde{k}} = y^{\alpha} \sum_{n=0}^{\infty} c_n y^n.
\end{equation}
The {\it indicial equation\/} is obtained by inserting this expression
into into Eq.(\ref{u_kmode_y}), expanding about $y=0$ and indentifying
the lowest order terms, which are proportional to $y^{\alpha-2}$.
This gives $\alpha =0$ or $3/2$.
Therefore, for small $y$,
\begin{equation}
u_{\tilde{k}} \sim c_{1} + c_{2} y^{3/2}.
\end{equation}
The coefficients $c_{1,2}$ are constrained by the Wronskian condition
which follows from the canonical commutation relation $\left[
\phi_{\tilde{k}}, \pi_{\tilde{r}} \right] = i \delta^3
(\tilde{k}-\tilde{r})$ for $\phi_{\tilde{k}}=u_{\tilde{k}}/a$ and its
conjugate momentum $\pi_{\tilde{r}}$ \cite{Kempf,KN} 
\footnote{In the model of \cite{Kempf}, the commutation
relations of ${\bf x}$ and ${\bf p}$ are modified as in
Eq.(\ref{uncertain}), but the field commutation relations are not
changed from their standard form. While one can seek a particular
interpretation, here we simply view this as providing an interesting,
short distance modification of the field equations that allows for
a reliable calculation of  cosmological implications.}:
\begin{equation}\label{wronskianu}
u_{\tilde{k}}(\eta) u_{\tilde{k}}^{* \prime} (\eta)- u_{\tilde{k}}^{*}
(\eta) u_{\tilde{k}}^{\prime} (\eta) = i \left(1-\beta \rho^2 \right)
\exp(-\frac{3}{2} \beta \rho^2).
\end{equation}
Hence,
\begin{equation}
 c_1 c_2^{*} - c_1^{*} c_{2} = i \eta_k \frac{4}{3} e^{-3/2}.
\end{equation}
The Wronskian condition mixes the two power series solutions with
complex coefficients.  {\em A priori}, there is a one parameter family
of solutions of $c_{1,2}$, corresponding to different choices of the
vacuum state.  In the usual inflationary scenario, the natural choice of
the vacuum state is the so-called Bunch-Davies vacuum which reduces to
the Minkowskian vacuum for wavelengths much shorter than the Hubble
scale.  We now motivate a natural choice of vacuum in the present
context by comparing with the Bunch-Davies vacuum. Recall that in the
standard setting, the high frequency limit of the mode equation takes
the form
\begin{equation}
u_{{k}}''(\eta) + \omega_k^2(\eta)u_{k}(\eta) = 0,
\label{uequation2}
\end{equation}
whose solution is well approximated by the WKB form
\begin{equation}
u_{{k}}^{\mp} (\eta) = \frac{1}{\sqrt{2 \omega_k}} \exp (\pm i
\int^{\eta} \omega_k(\eta') d \eta')
\end{equation}
if the adiabatic conditions $\frac{\omega_k^{\prime\prime}}{\omega_k}
\ll 1$ and $\left|\frac{\omega_k'}{\omega_k^2}\right| \ll 1$ are both
satisfied. The Bunch-Davies vacuum amounts to choosing the Heisenberg
form of the field operator ${\hat u}(\eta, {\bf x})$ to be
\begin{equation}
{\hat u}(\eta,{ \bf x}) = \int \frac{d^3k}{(2 \pi)^{3/2} } (u_k^+ a_k
e^{ikx} + u_k^{* +} a_k^{\dagger} e^{-ikx}), 
\end{equation}
with a vacuum state $|0\rangle$ satisfying $a_k |0\rangle = 0$.  In
our case, if we could ignore the $\dot u_{\tilde{k}}$ in
(\ref{u_kmode_y}), similar reasoning suggests
\begin{equation}
u^+_{\tilde{k}} (\eta) = \frac{1}{\sqrt{2 \omega_k}} \exp (-i \int^{\eta}
\omega_k(\eta') d \eta'),  
\end{equation}
where, from (\ref{u_kmode_y}), we see that for $\eta \sim \eta_k$,
\begin{equation}
\omega_k^2 = \frac{A}{\eta_k^2 y}
\end{equation} 
and
\begin{equation}
A =  \frac{1}{4 \beta H^2} + \frac{1}{2}.
\end{equation}
This would give a Bunch-Davies-like vacuum of the form
\begin{equation}
u^+_{\tilde{k}} (y) = 
\left( \frac{\eta_k^2 y}{4A} \right)^{1/4}
\exp(-2 i \sqrt{A y} ).
\end{equation} 
Of course, though, we can not ignore the $\dot u_{\tilde{k}}$ in
(\ref{u_kmode_y}), and hence this does not yield a solution to our
modified mode equation. However, by modifying the prefactor of the
exponential, we can in fact construct a solution of this form,
\begin{equation}\label{Fsoln}
F(y) =\left( \frac{\sqrt{A}}{2} + i A \sqrt{y} \right) \exp(-2i\sqrt{A y}).
\end{equation}
This solves Eq.~(\ref{u_kmode_y}); equivalently, in the series
expansion of $F(y)$ the $y^{1/2}$ term cancels out, and $F(y)$ is seen 
to be a linear combination of the two power series solutions found
above with $c_1 = \frac{\sqrt{A}}{2}$ and $c_2 = -i \frac{4}{3}A^2$.

The general solution is a linear combination of the positive and the
negative frequency modes
\begin{equation}
u_{\tilde{k}} (y) = C_{+} F(y) + C_{-} F^{*} (y)
\end{equation}
with the constants $C_{\pm}$ constrained by the Wronskian condition
\begin{equation}
|C_{+}|^2 - |C_{-}|^2 
= \frac{e^{-2}}{\sqrt{\beta} \tilde{k} H} 
\left( \frac{1}{4 \beta H^2} + \frac{1}{2} \right)^{-5/2}.
\end{equation}
Comparison of the osciallatory part of $F(y)$ with that of the
Bunch-Davies vacuum suggests that $C_{-}=0$.  However, the
normalization of $F(y)$ is not $1/\sqrt{2 \omega}$ since the adiabatic
condition is not satisfied. To be specific,
\begin{equation}
\frac{\omega_k^{\prime}}{\omega_k^2} = \frac{1}{2 \sqrt{A y}}.
\end{equation}
Therefore for $y \sim 0$, no matter how large $A$ is (or, more
importantly, how small $\beta$ is), the adiabatic condition
$|\frac{\omega_k^{\prime}}{\omega_k^2}| <<1$ is violated. Thus these
initial conditions are not smoothly connected to the ``standard'' form
of the mode equation when it is well inside the horizon, even in the
limit $\beta \rightarrow 0$ although the Lagrangian is smooth in the
same limit. The discrepancy arises because whenever $\beta \ne 0$, the
mode's evolution begins at a {\em finite\/} conformal time.  Moreover,
we expanded the Lambert function $W(z)$ with $z=-\beta
\tilde{k}^2/a^2$ around $z=-1/e$, and this expansion is only
convergent when $z<0$. If $\beta=0$, the argument of the $W$ function
is zero, and this expansion cannot be used.

We have found the leading behavior of $u_{\tilde{k}}$ around $\eta
\sim \eta_k$. The equation of motion for the tensor mode
$u_{\tilde{k}}$ is solved only up to order $1/y$.  The residual terms
can still be significant for $\eta \sim \eta_k$.  We deduce the
subleading bahavior of $u_{\tilde{k}}$ by the method of dominant
balance \cite{bender}.  Define $u_{\tilde{k}} (y)= F(y) \left(1 +
\epsilon_1 (y) \right)$ and extract the most singular terms in the
equation of motion for $\epsilon_1$:
\begin{equation}
\ddot{\epsilon}_1 - \frac{1}{2y}  \dot{\epsilon}_1 =
\frac{3A-3/2}{\sqrt{y}} = 0,
\end{equation}
which gives
\begin{equation}
\epsilon_1 (y) = 
(2A-1) y^{3/2}
\left( \log y -\frac{2}{3} \right).
\end{equation}
This is indeed small compared with the leading term for small $y$.
With this correction, the equation is solved up to order $1/\sqrt{y}$, but
there are still residual $\ln(y)$ terms. The solution is further improved
by the next subleading order $\epsilon_2 (y)$, where 
$u_{\tilde{k}}=F(y) (1+\epsilon_1(y)) (1+\epsilon_2 (y))$. The solution for
$\epsilon_2 (y)$ is
\begin{eqnarray}
\epsilon_2 &=& \frac{1}{24} \left( 105 - 330 A - 112 i
A^{3/2}\right)y^2 +
\nonumber \\
&& \quad \frac{7}{4}\left(2A -1 \right) y^2 \log{y}.
\end{eqnarray}
The equation of motion for $u_{\tilde{k}}$ is therefore solved up to
terms that vanish as $\eta \rightarrow 0$.

When the mode is well outside the horizon, $\rho \ll H$, and see that
$\mu(\eta,\rho) \rightarrow k^2$ and $\nu(\eta,\rho)
\rightarrow 1$. Consequently, $u_k (\eta) \sim a(\eta)$, which
reproduces the standard late time limit in the usual case with
$\beta=0$. This limit defines the power spectrum,
\be
{\cal P}_g (k) = \frac{k^3}{2 \pi^2} \left|\frac{u_k}{a}\right|^2.
\ee
which is evaluated when $u_k$ is well outside the horizon.

\begin{figure}[tb]
\begin{center}
\begin{tabular}{c}
\epsfxsize=8cm
\epsfbox{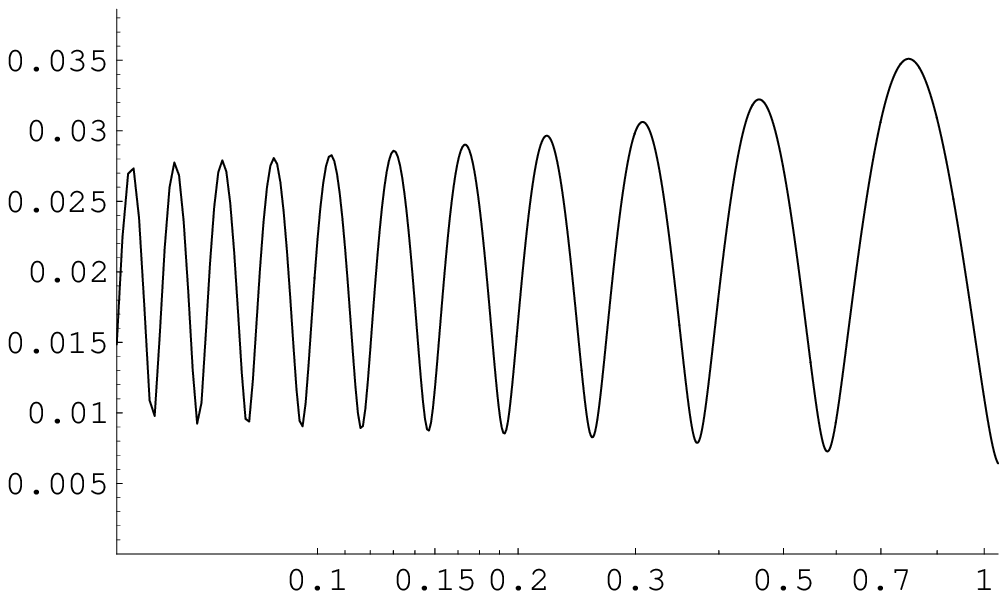} \\
\epsfxsize=8cm
\epsfbox{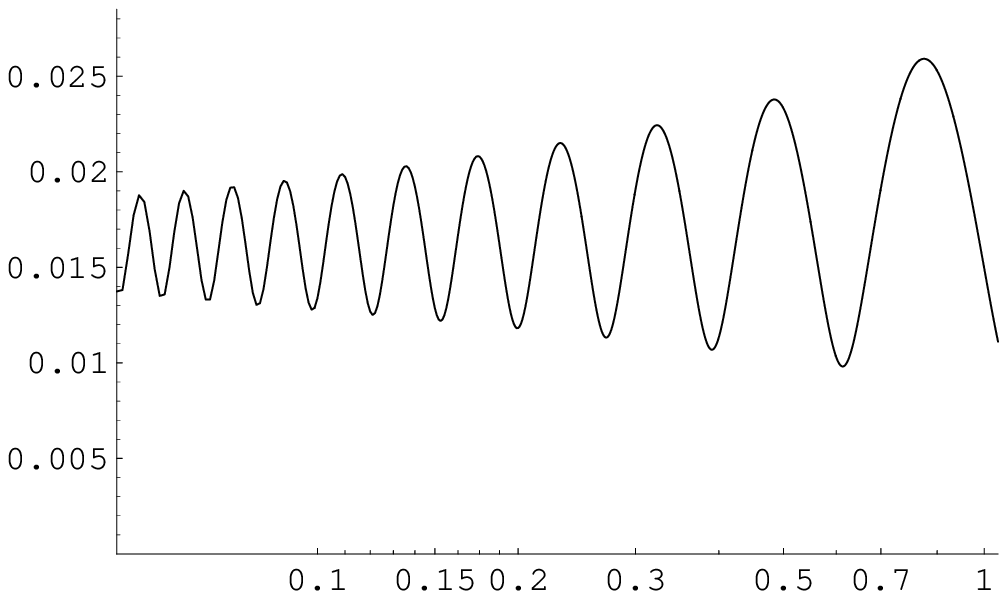}
\end{tabular}
\end{center}
\caption[]{ The top figure shows the spectrum ($P^{1/2}$) as a
function of $\beta$ with $C_{-}/C_{+} = -.5$, while the lower plot
shows the spectrum when $C_{-} = 0$. It is only in the latter case the
computed value of the spectrum smoothly approaches the usual value in
de Sitter space, In this plot $H=.1$, and with $\beta =0$ we would
expect $P^{1/2} = 0.0159155$.}
\label{mix}
\end{figure}

\begin{figure}[tb]
\begin{center}
\begin{tabular}{c}
\epsfxsize=8cm
\epsfbox{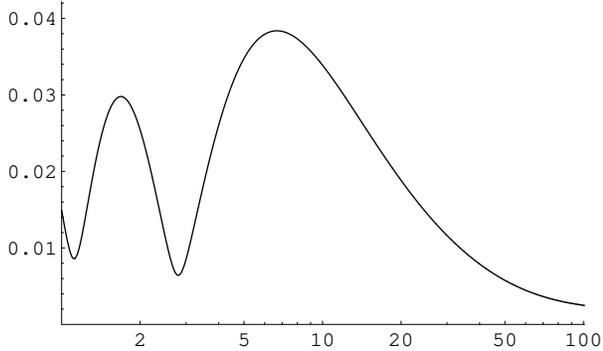} 
\end{tabular}
\end{center}
\caption[]{The $\beta$ dependence of the spectrum, $P_k^{1/2}$, is
plotted against $\beta$, with $H=0.1$.  For a de Sitter background,
the spectrum is independent of $k$. \label{spectrum}}
\end{figure}

\section{Numerical Results}

We match the analytical form of the solution valid when the mode is
well inside the horizon to a numerical evaluation of the full mode
equation.  In principle we might hope to evaluate the mode
equation numerically from an initial time of $y=0$, but the coefficients of
$u_{\tilde{k}}$ and $u_{\tilde{k}}'$ are infinite at this point so it
is easier to match the numerical solutions to the approximate
analytical solutions at $y_0$, a small but finite value of $y$. By
varying $y_0$ we can ensure that our results do not depend on our
choice of starting point, and that we are therefore ``close enough''
to $y=0$. 

A subtle point that arises during the integration is that the
subdominant terms in $u_{\tilde{k}}$ contribute terms of order
$y^{3/2} \log(y)$, which contribute $\sqrt{y} \log{y}$ in the
derivative, $u_{\tilde{k}}'$. While these terms do go to zero in the
limit where $y_0$ is vanishingly small, they approach zero
comparatively slowly. Since we are using a small but not infinitesimal
value of $y_0$, we match the numerical solution to the ``corrected''
asymptotic solution, which includes the lowest order logarithmic
terms. We evolve the mode equations numerically using the
Bulrisch-Stoer integrator implemented in Fortran, and from the
asymptotic values of $|u_k/a|$ we obtain the spectrum, $P^{1/2}(k)$.

In general the spectrum is $k$-dependent, but for the special case of
de Sitter inflation, the background spacetime is time translation
invariant, which implies that the spectrum should not depend on $k$
(which is equivalent to $\tilde{k}$ at late times). This is manifest
from the mode equation and, in practice, we verified the code by
solving the mode equation for multiple values of $k$ and found that
the numerically computed spectrum was indeed scale invariant to better
than 1 part in $10^6$.

In the previous section, we saw that imposing the Wronskian constraint
led to a one parameter family of solutions, but that a comparison with
the standard Bunch-Davies vacuum suggests the choice $C_{-} = 0$. We
begin by analysing the consequences of relaxing this choice. In
Fig.~(\ref{mix}) we plot the spectrum computed for small values of
$\beta$ with $C_{-} =0$, and compare this plot to a calculation with a
finite value of $C_{-}$.  In the former case, the spectrum smoothly
approaches the de Sitter result $P^{1/2} = H/2 \pi$, but does not do
so for any finite value of $C_{-}$.

Fig.~{\ref{spectrum}} displays the $\beta$ dependence of the spectrum.
For moderate values of $\beta$, the spectrum can be either lower or
higher than the ``standard'' value, but for large $\beta$, $P^{1/2}$
approaches zero.  For sufficiently large $\beta$, the overall
amplitude of the perturbation spectrum is significantly reduced. In de
Sitter space the power spectrum is $k$-independent, so the only effect
is an overall normalization dependence on $\beta$ through the
dimensionless combination ${\sqrt{\beta}} H$. Assuming one has an
independent measure of $H$, we see that the short distance cutoff
specified by $\beta$ does indeed leave an imprint on the spectrum.
In more realistic inflationary models, as we indicate later, we
believe the imprint will be $k$-dependent and hence also affect the shape
of the power spectrum.

\begin{figure}[tb]
\begin{center}
\begin{tabular}{c}
\epsfxsize=8cm
\epsfbox{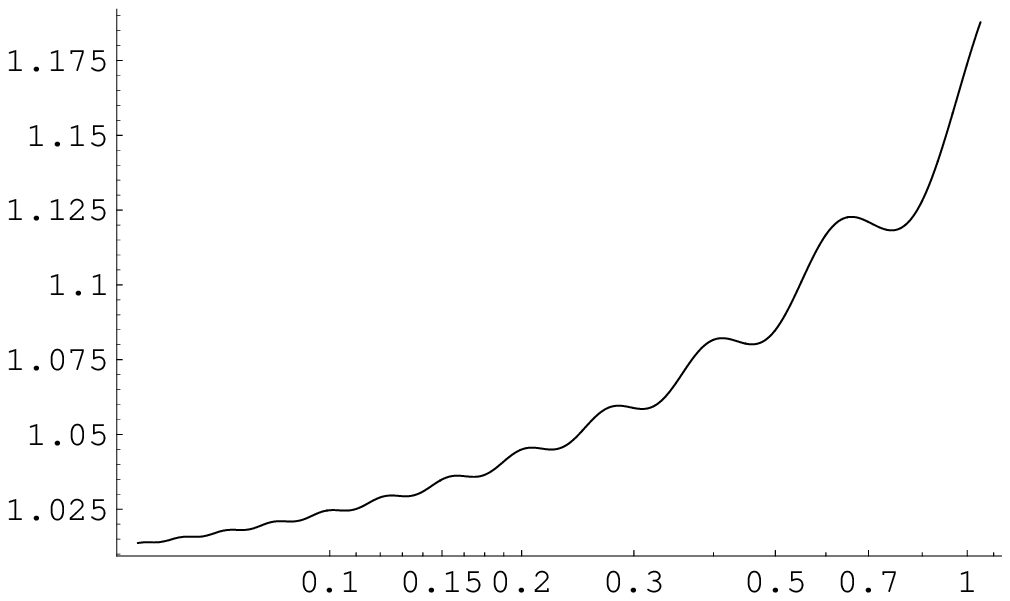} \\
\epsfxsize=8cm
\epsfbox{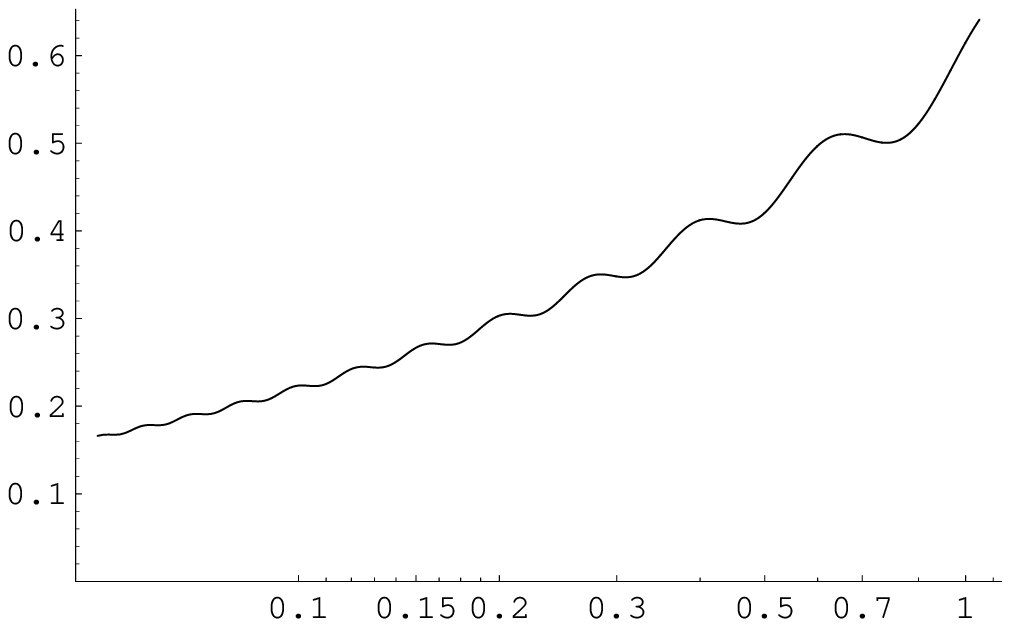}
\end{tabular}
\end{center}
\caption[]{The values of $D_{+}$ (top) and $D_{-}$ are extracted for
the spectrum shown in Fig.~1 with $C_{-} =0$. As $\beta \rightarrow
0$,  $D_{+} \rightarrow 1$ and $D_{-} \rightarrow 0$, confirming that as 
the Hubble length becomes much larger than the minimum length the
spectrum approaches the standard result. }
\label{Dvals}
\end{figure}

\section{Mode Matching After Horizon Crossing}

The physical origin of the change in normalization of the power
spectrum can be simply understood in terms of the asymptotic behavior
of the mode solutions. In the long wavelength limit, the mode equation
asymptotes to the standard mode equation for cosmological
perturbations.  Therefore, regardless of the nature of the
short-distance physics, the mode function takes the general form:
\begin{equation} \label{longlimit}
u_k\left(\eta\right) = \frac{1}{2}\sqrt{-{\pi \eta}} \left[D_{+}
H_\nu\left(-k \eta\right) + D_{-} H^*_\nu\left(- k \eta\right)\right],
\end{equation}
where $H_\nu$ is a Hankel function of the first kind, and $v = 3/2$ in
de Sitter space. The constants $D_{\pm}$ are constrained by the
standard Wronskian condition arising from the canonical commutation
relations for the field operator:
\begin{equation}
u_k^* {d u_k \over d \eta} - u_k {d u_k^* \over d \eta} = - i.
\end{equation}
With this constraint, the form of the mode function is completely
determined (up to a phase) by the selection of a vacuum. The standard
case, that of a Bunch-Davies vacuum at short wavelength, corresponds
to the selection $D_{-} = 0$ and $D_{+} =1$, with our choice of
normalization. 

The key attribute of this construction is that {\em regardless} of the
nature of the short-distance physics, the long-wavelength behavior of
the modes is completely encoded in the coefficients $D_{\pm}$. In
general the $D_{\pm}$ are $k$-dependent but in de Sitter space the
situation is particularly simple, since the $D_{\pm}$ are 
constants.

In the standard analysis, the coefficients $C_{\pm}$ that determine
the vacuum at short wavelengths are identical to the $D_{\pm}$ found
by matching the mode functions to the long-wavelength limit,
Eq.~(\ref{longlimit}).  Modifying the short distance physics can break
this correspondence, and the choice of vacuum $C_{-} = 0$ in the short
wavelength limit can correspond to a phase rotation in the long
wavelength limit, with $D_{-} \neq 0$. This changes the normalization
of $u_k$, since the power spectrum depends on the integration
constants as
\begin{equation}
{\cal P}_g (k) = \frac{k^3}{2 \pi^2} \left|\frac{u_k}{a}\right|^2 
   \propto \left|D_{+} + D_{-}\right|^2
\end{equation}
in the long wavelength ($k \rightarrow 0$) limit.  In de Sitter space,
the $D_{\pm}$ are constant, so the rotation alters the normalization
of the power spectrum, but does not introduce any $k$ dependence.
However, if the background deviates from de Sitter the $D_{\pm}$ are,
in general, $k$-dependent, which will alter both the shape and
normalization of the power spectrum as well as the short distance
physics -- perhaps in a dramatic way \cite{EGKS}.  Naturally, if a
modification to short-distance physics alters $D_{\pm}$, the
perturbation spectrum will change.  This will lead to constraints on
the form and magnitude of the modifications to $D_{\pm}$.

\section{Particle Production Bounds}

In addition to the perturbation spectrum, other arguments place
cosmological constraints on modifications to short-distance
physics. For example, Starobinsky\cite{starobinsky} has recently shown
on quite general grounds that even a very tiny rotation away from the
Bunch-Davies vacuum in {\em the current universe\/} would result in
unacceptable production of relativistic particles. The Starobinsky
bound corresponds to
\begin{equation}
\left|D_{-}\right|^2 \leq \frac{H_0^2}{M_{\rm Pl}^2},
\end{equation}
where the coefficient $D_{-}$ is evaluated in the limit of a scale
much larger than the minimum distance $\sqrt{\beta}$ but still much
smaller than the current horizon size $H_0^{-1}$. However, this bound
applies to the cosmic vacuum at late times, not during inflation. If
$\left|D_{-}\right|$ is dependent on $H$, it can be significant during
inflation, thus affecting the power spectrum, but small enough
universe today to satisfy the Starobinsky bound.

As it happens, an $H$ dependent value of $D_{-}$ is precisely what the
model considered in this paper predicts. The magnitude of the rotation
depends not on the absolute physical scale of the cuttoff, but on the
ratio of the cutoff scale to the horizon size, $\sqrt{\beta} H$. If
the horizon size and the minimum length are comparable during
inflation, the power spectrum is significantly affected, but the rate
of particle production today will be strongly suppressed, by the much
lower value of $H$ in the present universe.  Fig. 3 shows the
dependence of the coefficient $D_{-}$ on the ratio $\beta H^2$ in the
numerical solution of the mode equation.  The numerical results imply
$\left|D_{-}\right| \propto \beta^n$, with $0.45 \lesssim n \le 0.5$
when $\beta H^2 \ll 1$.  In the ``real'' inflationary universe, we
hold $\beta$ fixed and reduce $H$, so this result implies that as the
universe expands the $D_{+}$ term will become negligible.

On the basis of our present calculations, the exact dependence of
$D_{-}$ on $\beta$ remains unclear, since the value of $n$ appears to
be very weakly dependent on $\beta$: if we evaluate $n$ over a few
decades in $\beta$ (for fixed $H$) $n$ appears to slowly approach
$0.5$ as this range is moved to smaller and smaller $\beta$. Thus, it
is not unreasonable to assume that $\left|D_{-}\right| \propto
\beta^{.5}$ in the present universe, and that the particle production
rate today is consistent with the Starobinsky bound.  

We do not pursue this in more detail as the current model is simply a
heuristic construction that mimics what we might expect from a more
rigorous string theoretic description of spacetime.  However, we can
conclude that it is plausible that a minimum length leads to a time
dependent modification of the backgroun in a universe where $H$ is not
constant, and that the value of $D_{-}$ will be time dependent making
the particule production bound much less onerous.

Also of relevance is Tanaka's analysis \cite{tanaka00}, showing that
particle production during inflation can significantly change the
background evolution of the spacetime. The perturbation to the
stress-energy during inflation can be expressed as:
\begin{eqnarray}
\delta T^{00} &&\sim \int{{d^3 k \over \left(2 \pi\right)^3} 
 \left[{k \left|D_{-}\right|^2 \over a^4}\right]}\cr
&&= \int{{d^3 p \over \left(2 \pi\right)^3} p \left|D_{-}\right|^2}\cr
&&\sim {\left|D_{-}\right|^2 \over \beta^2},
\end{eqnarray}
where $p$ is the physical momentum $p \equiv \left(k / a\right)$, with
the momentum cutoff $p < \left(1 / \sqrt{\beta}\right)$. If
perturbation theory is to be consistent, the contribution from the
stress-energy from particle production must be subdominant,
\begin{equation}
\delta T^{00} \sim {\left|D_{-}\right|^2 \over \beta^2} <<  M_{\rm Pl}^2 H^2.
\end{equation}
Tanaka's analysis assumed $\left|D_{-}\right| \sim 1$, resulting in a
bound on $\beta$ of
\begin{equation}
{1 \over \beta} << \sqrt{M_{\rm Pl} H^2}.
\end{equation}
In this model, $D_{-}$ is $\beta$-dependent, weakening of the Tanaka
bound, and for $D_{-} \sim \sqrt{\beta} H$ it becomes
\begin{equation}
{1 \over \beta} << M_{\rm Pl}^2.
\end{equation}
This is easily satisfied if $\sqrt \beta$ is within an order of
magnitude or two of the Planck length.  Therefore particle production
places no significant constraints on the viability of this model.

\section{Discussion}

We have presented an accurate calculation of the perturbation spectrum
predicted for a de Sitter universe where the physics has been modified
to include a minimum length.  The spectrum is scale independent, which
follows from the time translation invariance of a de Sitter universe.
This calculation significantly extends the contemporaneous
qualitative approach of \cite{KN}, allowing us to determine {\it how}
the perturbation spectrum is modified.  Furthermore, it lays the
groundwork for accurate calculations of the spectrum produced by other
models where trans-Planckian physics has modified the ``standard''
evolution of cosmological perturbations, either by altering the
dispersion relationship in the evolution equations, or by changing the
initial conditions.

If the minimum length is much smaller than the Hubble length, its
introduction has no detectable effect on the spectrum.  However if
these two lengths are within an order of magnitude or two, it is possible for
the resulting spectrum to differ appreciably from the $\beta=0$
limit and for various other particle production constraints,
discussed above, to be satisfied.
Various values for the string scale and the Hubble scale arise
in recent string theoretic approaches to cosmology 
(see for example
\cite{add,ST,BW,ovrut,RS,dvali-tye,savas})
and hence the effects studied here can potentially be significant in
these and other models.

In de Sitter space, the ratio between the minimum length and the
physical horizon size is constant.  In almost all other inflationary
backgrounds, the expansion rate is slower than exponential, and the
physical horizon size will increase relative to the minimum length
scale. In this case our analysis of the de Sitter background suggests
that the amplitude of the longest modes (produced earliest in
inflation) will be modified. Shorter modes will leave the horizon at a
time when the horizon length is much larger than the physical cut-off
length and their amplitude will be unaffected.  We plan to return to
this problem in future work, but our tentative conclusion is that the
spectrum of primordial perturbations could be altered at very long
wavelengths by the existence of a minimum length scale. Whether this
is observable in practice will depend crucially on the number of
e-folds of inflation preceding the creation of the modes which are
responsible for large scale structure in the observable universe.

We close with a few observations, some rather speculative, about
further studies we plan to undertake based on the present work
\cite{EGKS}.

$\bullet$ All of the calculations in this paper have focused on
tensor perturbations. It would be worthwhile to extend the analysis to
scalar perturbations.

$\bullet$ We find it particularly interesting that since short scale
modifications yield mode equations that are asymptotic to the standard
form on large scales (that is what is meant by a short scale
modification) the effect on the power spectrum of {\it any} new short
distance physics can be encoded in the $k$-dependent coefficients
$D_+$ and $D_-$. This provides a phenomenological framework for
systematically seeking signals of -- and establishing constraints on
-- short scale deviations from conventional physics.

$\bullet$ For very large $\beta$, the power spectrum
becomes arbitrarily small. It is tempting to interpret this as a
mechanism for solving the fine tuning problem endemic to the
inflationary generation of the primordial perturbation spectrum. A
mechanism ensuring that the overall normalization of the spectrum is
small enough to satisfy observational constraints from Large Scale
Structure and the microwave background will solve this problem. For
large $\beta$, this model appears to do just that, but we caution that
this conclusion is likely na{\"\i}ve since it requires the minimum
physical length to be much larger than the horizon volume, and we can
not be sure the model has any physical validity in this regime. Further study
along these lines may reveal a trustworthy suppression mechanism.

$\bullet$ An interesting -- but highly speculative -- possibility is that
particle production induced by a finite value of $D_{-}$ may both be
small enough to satisfy Starobinsky's bound, but large enough to lead
to new physics. Epochs where this would be particularly
intriguing are immediately after inflation, where particle production
from the vacuum is a potential mechanism for reheating the universe,
and in the present universe if the particle production was efficient
enough to modify the equation of state. To explain the latter
possibility, the equation of state determines the relationship between
the density and the scale factor as the universe expands and particle
production reduces the rate at which the density of the universe drops
with increasing volume. In general, the more weakly the density of a
perfect fluid depends on volume, the more rapidly the universe will
expand. Consequently, if particle production is efficient enough to
alter the expansion rate of the universe, it will be increased
relative to that of a universe with $D_{-}=0$. This is particularly
interesting in the light of observational evidence for dark energy,
which is betrayed by a too-rapid expansion of the spacetime
background. For an alternative approach of understanding the origin
of dark energy
from trans-Planckian physics, see \cite{Mersini}.

\section*{Acknowledgments}
R.E. and W.H.K. are supported by the Columbia University Academic
Quality Fund and the Department of Energy.  The work of B.R.G. is
supported in part by the DOE grant DE-FG02-92ER40699B.  The work of
G.S. was supported in part by the DOE grant DE-FG02-95ER40893 and the
University of Pennsylvania School of Arts and Sciences Dean's fund.
We thank Philip Argyres, Vijay Balasubramanian, Clifford Burgess,
Chong-Sun Chu, Eanna Flanagan, Achim Kempf, Jens Niemeyer, Lam Hui and
Henry Tye for discussions.

\end{document}